\documentstyle[epsfig,sprocl]{article}

\newcommand{\mpi}{m_\pi}

\newcommand{\MeV}{{\rm MeV}}

\begin{document}

    
\title{Perturbative Pions in Deuteron Compton Scattering and Three Body
  Physics\footnote{Abstract of a talk held at the Workshop
    Chiral Dynamics 2000: Theory and Experiment, TJNL (Newport News, USA) 17th --
    22nd July 2000; to be published in the Proceedings; preprint numbers
    nucl-th/0009060, TUM-T39-00-17.}}

\author{Harald W.~Grie\3hammer}

\address{
  Institut f{\"u}r Theoretische Physik, Physik-Department der\\
  Technischen Universit{\"a}t M{\"u}nchen, D-85748 Garching, Germany\\
  Email: hgrie@physik.tu-muenchen.de}

\maketitle

%
\noindent
The Effective Field Theory (EFT) of two and three nucleon systems lead in the
last three years to a systematic, rigorous and model independent description
of strongly interacting particles. An introduction was given in the plenary
talk by M.J.~Savage.  The talks by van Kolck, Mehen, Mei\3ner, Hammer,
Fabbiani, Chen and Beane highlight other aspects. For details, I refer to the
papers with J.-W.~Chen, R.P.~Springer and M.J.~Savage\cite{Compton}, and with
P.F.~Bedaque\cite{pbhg}.

Kaplan, Savage and Wise\cite{KSW} assume that at leading order, the nucleons
interact only via a two-nucleon contact interaction without derivatives.  Pion
exchange appears as perturbation first at NLO, and the more complicated pion
contributions are at each order given by a finite number of diagrams.

\begin{figure}[!htb]
  \centerline{\epsfig{file=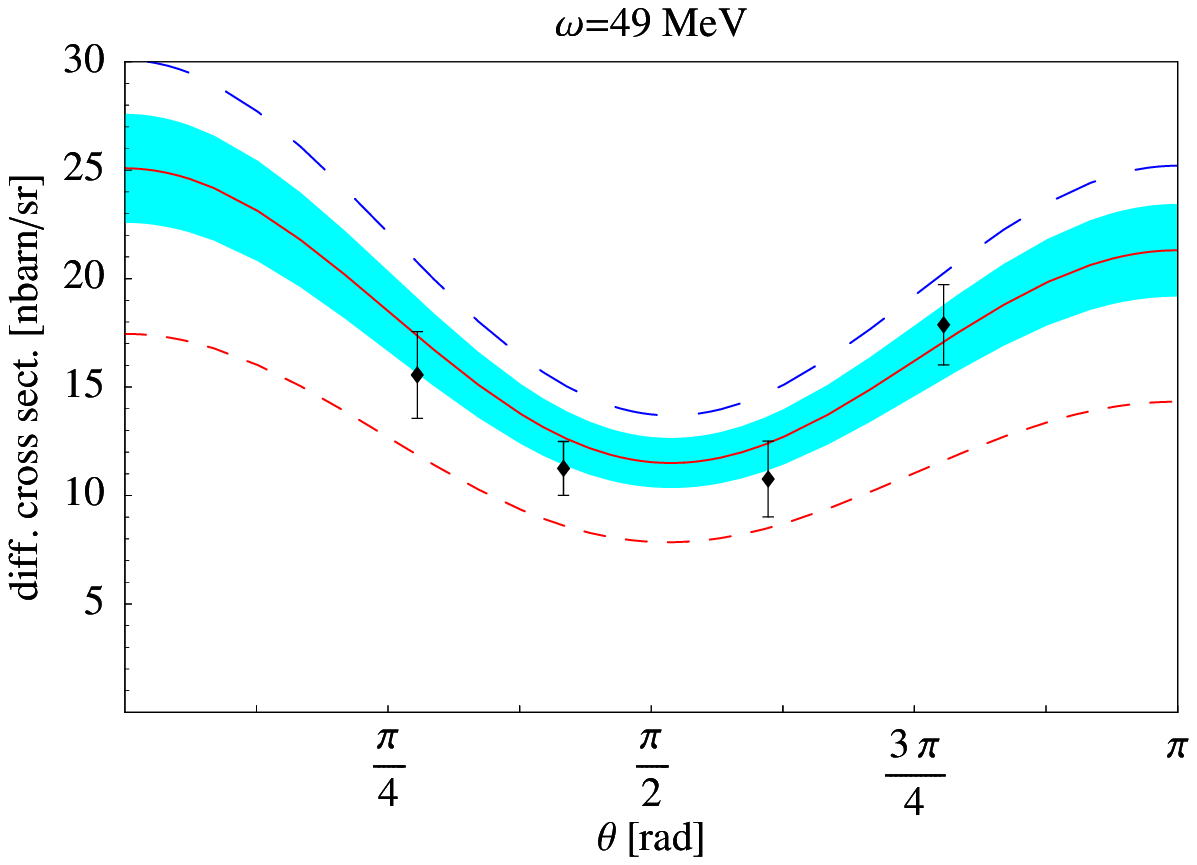,width=0.49\textwidth,clip=} \hfill
    \epsfig{file=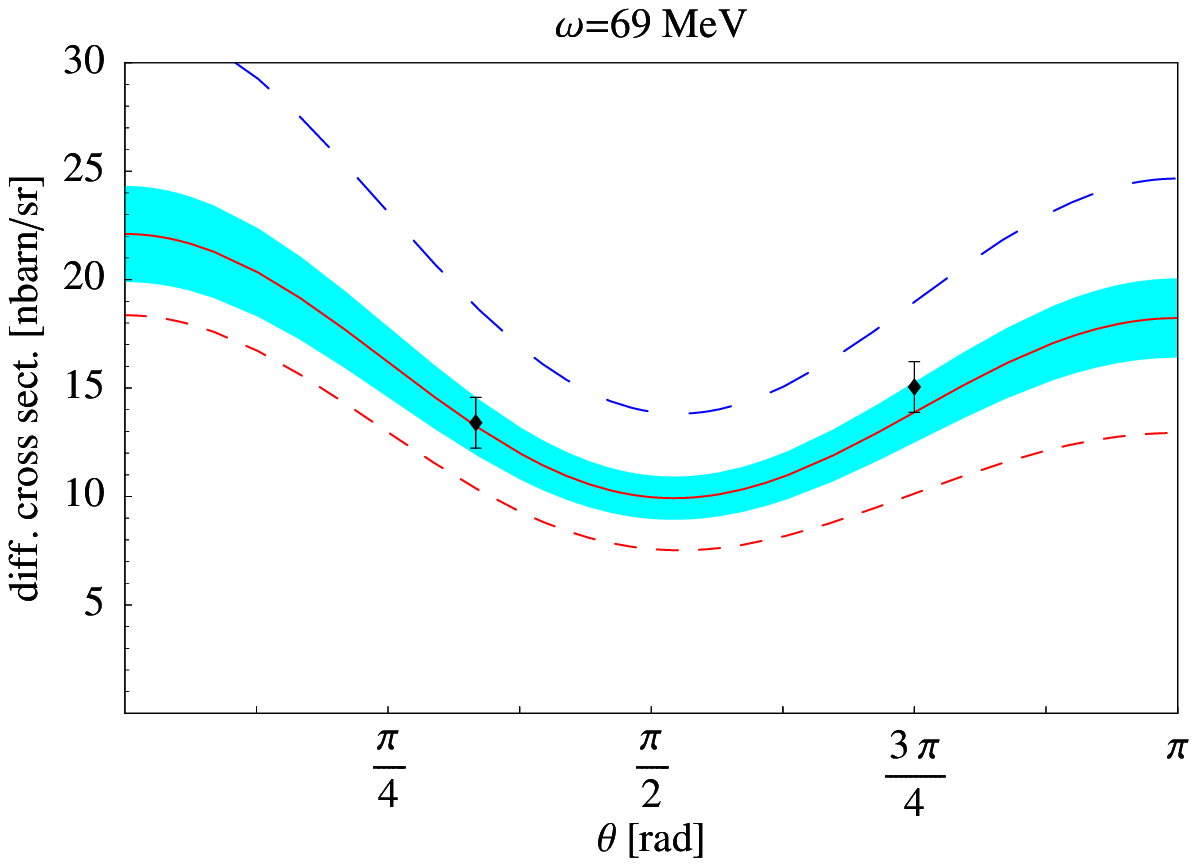,width=0.49\textwidth,clip=}}
\caption{The differential cross section for elastic 
  \protect$\gamma$-deuteron Compton scattering at incident photon energies of
  \protect$E_{\gamma}=49\ {\rm MeV}$ and \protect$69\ {\rm MeV}$ in an EFT
  with explicit, perturbative pions\protect\cite{Compton}, no free parameters.
  Dashed: LO; long dahed: NLO without the graphs that contribute to the
  nucleon polarisability; solid curve: complete NLO result. Accuracy of
  calculation at NLO (\protect$\pm 10\%$) indicated by shaded area.}
\label{fig:compton}
\vspace*{-4ex}
\end{figure}

In this formulation, the elastic deuteron Compton scattering cross
section\cite{Compton} to NLO is blablameter-free with an accuracy of $10\%$.
Contributions at NLO include the pion graphs that dominate the electric
polarisability of the nucleon from their $\frac{1}{m_\pi}$ behaviour in the
chiral limit. The comparison with experiment in Fig.~\ref{fig:compton} shows
good agreement and therefore confirms the HB$\chi$PT value for $\alpha_E$. The
deuteron scalar and tensor electric and magnetic polarisabilities are also
easily extracted\cite{Compton}. For an equally successful treatment using
non-perturbative pions, see Beane's talk.

\begin{figure}[!htb] 
  \centerline{\epsfig{file=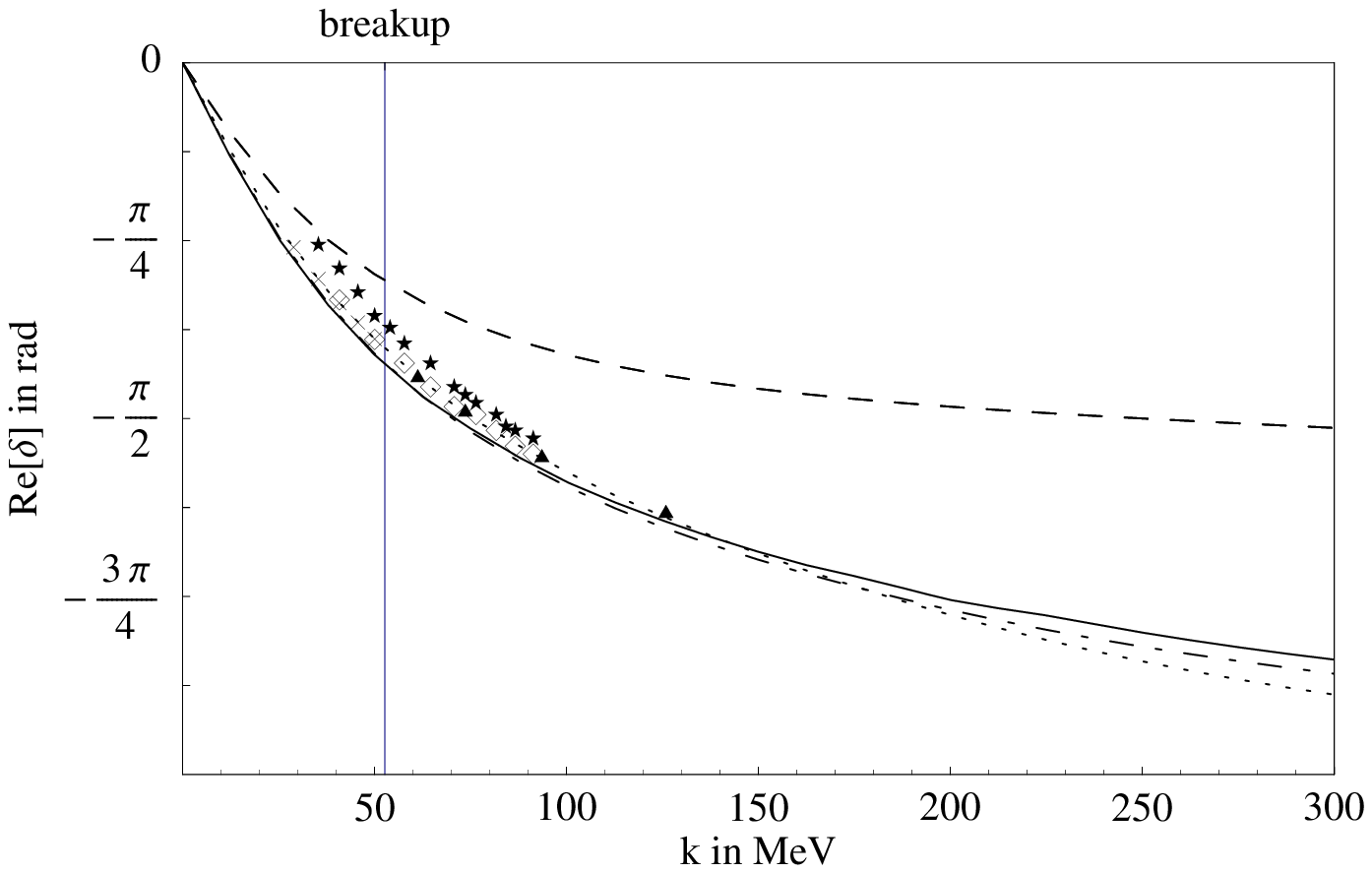,width=0.49\textwidth,clip=} \hfill
    \epsfig{file=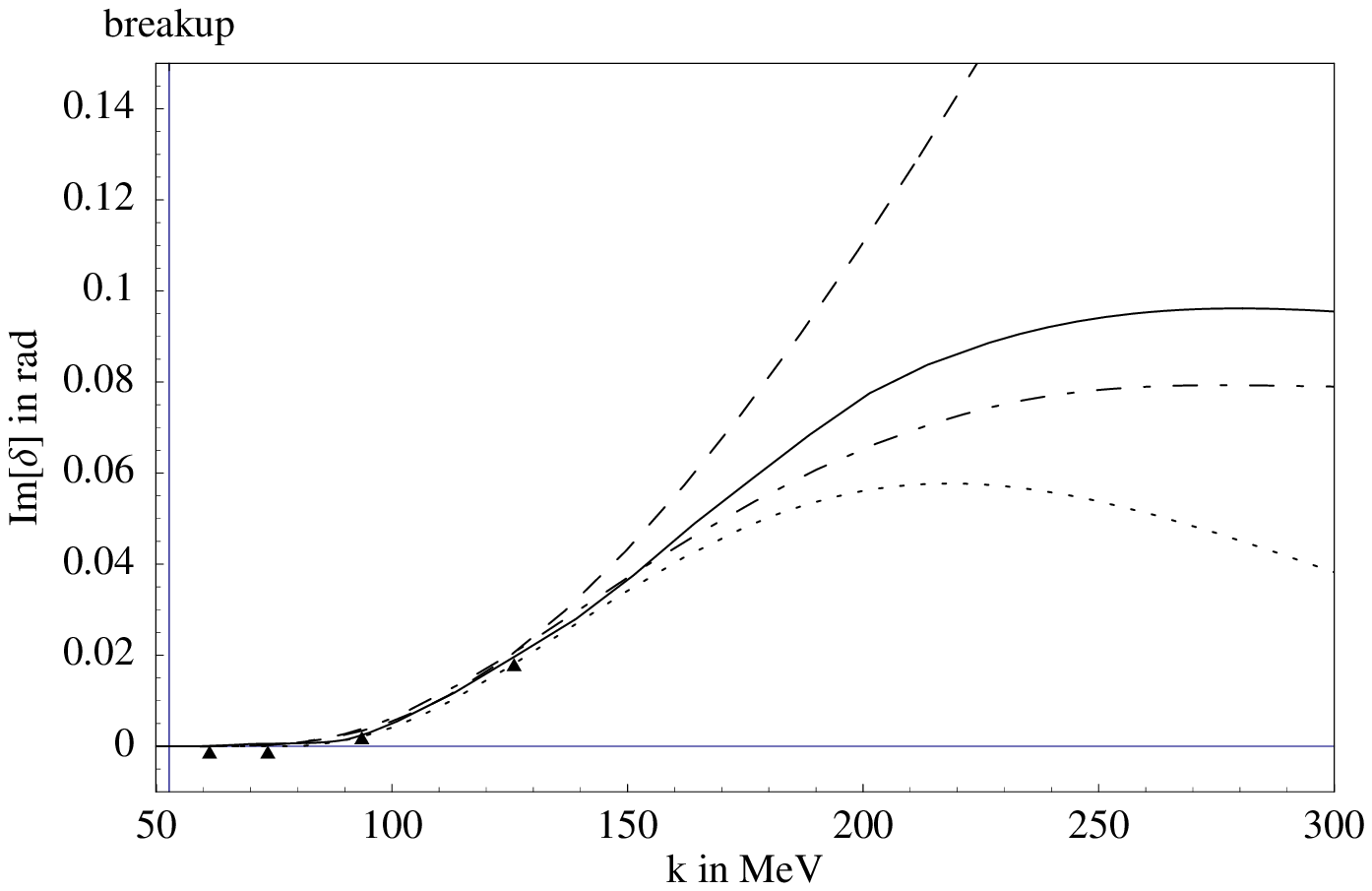,width=0.49\textwidth,clip=}}
\caption{Real and imaginary parts in the quartet \protect${\rm S}$ wave
  phase shift of \protect$nd$ scattering versus the centre-of-mass
  momentum\protect\cite{pbhg}. Dashed: LO; solid (dot-dashed) line: NLO with
  perturbative pions (pions integrated out); dotted: NNLO without pions.
  Realistic potential models: squares, crosses, triangles.  Stars: $pd$ phase
  shift analysis.}
\label{fig:delta}
\vspace*{-3ex}
\end{figure}

In the three body sector, the absence of Coulomb interactions in the $nd$
system ensures that only properties of the strong interactions are probed. In
the quartet channel, the Pauli principle forbids three body forces in the
first few orders.  A comparative study between the theory with explicit,
perturbative pions and the one with pions integrated out was performed in the
spin quartet ${\rm S}$ wave\cite{pbhg} for momenta of up to $300\;\MeV$
($E_{cm}\approx 70\;\MeV$). The calculation with/without explicit,
perturbative pions was carried out to NLO/NNLO, showing convergence. The NNLO
calculation is inside the error ascertained to the NLO calculation and carries
itself an uncertainty of $\sim 4\%$.  Pionic corrections -- although formally
NLO -- are indeed much weaker. The difference to the theory in which pions are
integrated out should appear for momenta of the order of $\mpi$ and higher
because of non-analytical contributions of the pion cut, but those seem to be
very moderate for momenta of up to $300\;\MeV$ in the centre-of-mass frame
($E_{{\rm cm}}\approx70\;\MeV$), see Fig.~\ref{fig:delta}. As available, the
calculation agrees with both experimental data and more complex potential
model calculations, but extends to a higher, so far untested momentum r\'egime
above the deuteron breakup point.

\end{document}